\def\slash#1{\setbox0=\hbox{$#1$}#1\hskip-\wd0\dimen0=5pt\advance
       \dimen0 by-\ht0\advance\dimen0 by\dp0\lower0.5\dimen0\hbox
         to\wd0{\hss\sl/\/\hss}}
\documentstyle[12pt]{article}

\newlength{\dinwidth}
\newlength{\dinmargin}
\newcommand{\resection}[1]{\setcounter{equation}{0}\section{#1}}

\begin{document}

\def\lq{\left [}
\def\rq{\right ]}
\def\LL{{\cal L}}
\def\VV{{\cal V}}
\def\AA{{\cal A}}
\def\gi{{g_{P^* P \pi}}}
\def\gis{{g_{P^* P^* \pi}}}
\def\lq{\left [}
\def\rq{\right ]}
\def\qq{<{\overline u}u>}
\def\dmu{\partial_{\mu}}
\def\dmus{\partial^{\mu}}

\def\gid{{g_{D^* D \pi}}}
\def\gib{{g_{B^* B \pi}}}
\def\gids{{g_{D^* D^* \pi}}}
\def\gibs{{g_{B^* B^* \pi}}}
\def\mbs{{m_{B^*}}}
\def\fbs{{f_{B^*}}}

\newcommand{\be}{\begin{equation}}
\newcommand{\ee}{\end{equation}}
\newcommand{\bea}{\begin{eqnarray}}
\newcommand{\eea}{\end{eqnarray}}
\newcommand{\nn}{\nonumber}
\newcommand{\dd}{\displaystyle}
\vspace*{0.8cm}
\begin{center}
  \begin{Large}
  \begin{bf}
HEAVY MESON HYPERFINE SPLITTING: A COMPLETE $1/m_Q$ CALCULATION.
\footnote{Partially supported by the Swiss National Foundation.}\\
  \end{bf}
  \end{Large}
  \vspace{5mm}
  \begin{large}
N. Di Bartolomeo \\
  \end{large}
{\it D\'epartement de Physique Th\'eorique, Univ. de Gen\`eve}\\
\vspace{5mm}
\end{center}
\begin{quotation}
\vspace*{1cm}
\begin{center}
  \begin{large}
  \begin{bf}
  ABSTRACT
  \end{bf}
  \end{large}
\end{center}
  \vspace{5mm}
\noindent
We compute the chiral corrections to the hyperfine splittings
$\Delta_D=(m_{D^*_s}-m_{D_s})-(m_{D^{*+}}-m_{D^+})$
and
$\Delta_B=(m_{B^*_s}-m_{B_s})-(m_{B^{*0}}-m_{B^0})$
arising from  one-loop chiral corrections, working in a framework
of an effective chiral lagrangian   incorporating
 chiral, heavy flavour and spin symmetric terms and
first order breaking terms.
Among these terms, those responsible for
the spin-breaking difference between the couplings $\gis$ and $\gi$ are
evaluated in the QCD sum rules approach.
Their contribution to $\Delta_D$ and to $\Delta_B$ appears to
cancel  previously estimated large chiral effects,
giving an estimate in agreement with the experimental data.
\end{quotation}
\vspace{3cm}
\begin{center}
{\it to appear in the Proceedings of the XXXth Rencontres de Moriond\\
``QCD and High Energy Hadronic Interactions'' \\
Les Arcs, France, March 1995 }
\end{center}

\newpage
\setcounter{page}{1}
\resection{Introduction}
The spectroscopy of heavy mesons is among the simplest framework
where the ideas and the
methods of heavy quark expansion can be quantitatively tested.
Recently, attention has been focused on
the combinations \cite{ros,ran,jen,chi}:
\be
\Delta_D=(m_{D^*_s}-m_{D_s})-(m_{D^{*+}}-m_{D^+})
\label{a}
\ee
\be
\Delta_B=(m_{B^*_s}-m_{B_s})-(m_{B^{*0}}-m_{B^0})
\label{b}
\ee
which are measured to be \cite{PDB}:
\be
\Delta_D\simeq 1.0\pm1.8~MeV
\label{c}
\ee
\be
\Delta_B\simeq 1.0\pm2.7~MeV
\label{d}
\ee
The above hyperfine splitting is free from electromagnetic corrections
and it vanishes
separately in the $SU(3)$ chiral limit and in the heavy quark limit.
In the combined
chiral and heavy quark expansion, the leading contribution is
of order $m_s/m_Q$ and
one would expect the relation \cite{ros}:
\be
\Delta_B=\frac{m_c}{m_b}\Delta_D
\label{e}
\ee
In the so called heavy meson effective theory \cite{wis}, which combines
the heavy quark expansion and the chiral symmetry, there is only one
lowest order operator
contributing to $\Delta_{D,B}$. By naive dimensional analysis,
its contribution to the
hyperfine splitting is of the order
\be
\Delta^{(2)}_D\simeq 20~MeV
\label{h}
\ee
\be
\Delta^{(2)}_B\simeq 6~MeV
\label{i}
\ee
Given the present experimental accuracy,
the above estimate is barely acceptable, as an order of magnitude,
for $\Delta_B$,
while it clearly fails to reproduce the data for $\Delta_D$.

In chiral perturbation theory, an independent contribution arises from
one-loop corrections to the heavy meson self energies \cite{jen},
evaluated from an initial lagrangian containing, at the lowest order,
both chiral breaking and spin breaking terms. The loop corrections
in turn depend on an arbitrary renormalization point $\mu^2$ (e.g. the
t'Hooft mass of dimensional regularization). This dependence is cancelled by
the $\mu^2$ dependence of a counterterm.
 A commonly accepted
point of view is that the overall effect of adding the counterterm
consists in replacing $\mu^2$ in the loop corrections with
the physical scale relevant to the
problem at hand, $\Lambda_{CSB}^2$. Possible finite terms in the counterterm
are supposed to be small compared to the large chiral logarithms.
With this philosophy in mind,
two classes of such corrections has been estimated in ref. \cite{ran}:
keeping the chiral logarithms  and non-analytic contributions of the order
$m_s^{3/2}$, they found a quite large correction to the hyperfine splitting,
\be
\Delta^0_D\simeq + 95~MeV,
\label{l}
\ee
\be
\Delta^0_B\simeq + 32~MeV,
\label{m}
\ee
This provides a rather uncomfortable situation since, to account for the
observed data, one should require an accurate and
innatural cancellation.

Hower, as pointed out in \cite{chi}, there is another term induced
 by  the difference between the $P^* P^* \pi$
and the $P^* P \pi$ couplings ($P=D,B$).
They coincide in the limit $M_P \to \infty$
\be
\gi=\gis=g
\label{equal}
\ee
because their splitting is a spin breaking effect.
To the order $1/m_Q$ we parametrize
them in the following way:
\be
\gis= g \left( 1+\frac{a}{m_Q} \right) \; \;\;\;~~
\gi= g \left( 1+\frac{b}{m_Q} \right)  \label{17b}
\ee

The chiral and spin symmetry breaking parameters relevant
to the hyperfine splitting
are the light pseudoscalar masses $m_\pi$, $m_K$ and $m_\eta$,
$\Delta_s = M_{P_s} - M_P$, $\Delta = M_{P^*} - M_P$ and
$\Delta_g = \gis -\gi$.
In terms of these quantities, one finds \cite{jen,ran,chi}:
\bea
\Delta_{P}&=&\frac{g^2 \Delta}{16 \pi^2 f^2}
            \Bigl[4 m_K^2 ln(\frac{\Lambda_{CSB}^2}{m_k^2})+
                  2 m_\eta^2 ln(\frac{\Lambda_{CSB}^2}{m_\eta^2})-
                  6 m_\pi^2 ln(\frac{\Lambda_{CSB}^2}{m_\pi^2})\Bigr]\nn\\
          &+&\frac{g^2 \Delta}{16 \pi^2 f^2} [24 \pi m_K \Delta_s]\nn\\
          &-&\frac{g^2}{6 \pi f^2}\frac{\Delta_g}{g}
             (m_K^3+\frac{1}{2} m_\eta^3-\frac{3}{2} m_\pi^3)
\label{split}
\eea
The dependence upon the heavy flavour $P=D,B$ is
contained in the parameters $\Delta$
and $\Delta_g$.

In \cite{hyperspl} we have  provided an estimate of $\Delta_g =
g_{P^* P^* \pi}-g_{P^* P \pi}$
based on a QCD sum rule, and, by including this additional spin breaking
effect,
we have  completed the evaluation of $\Delta_{D,B}$ in (\ref{split}).

\resection{ QCD Sum Rules for $\gi$ and $\gis$}

The coupling $\gi$ has  been calculated in \cite{noi},
$\gis$ in \cite{hyperspl},
by means of QCD
sum rules \cite{gensum}.
Without entering into the details of the calculation, we sketch
the strategy  followed in computing $\gis$ \cite{hyperspl}.

One starts  from the correlator:
\be
A_{\mu\nu}(q_1,q) = i \int dx <\pi(q)| T(V_{\mu}(x) V_{\nu}^{\dagger}(0)
 |0> e^{-iq_1x} = A(q_1^2, q_2^2,q^2)\epsilon_{\mu\nu\alpha\beta} q^{\alpha}
q_1^{\beta}+ \ldots
\label{2}
\ee
where $V_{\mu}={\overline u} \gamma_{\mu} Q$  is the interpolating vector
current for the $P^*$ meson,
 computing the scalar function $A$ in the soft pion limit $q \to 0$ (this
implies $q_1=q_2$ forcing to use a single Borel transformation).
The correlator in (\ref{2})
can be calculated by an Operator Product Expansion: in \cite{hyperspl}
 all the operators
with dimension up to five, arising from the expansion of the current
$V_{\mu}(x)$ at the third order in powers of $x$ and the heavy quark propagator
to the second order, are kept.

Proceeding in a standard way, one computes  the hadronic side of the sum
rule, equating it to the QCD side. The Borel transform enhances the
ground state contribution to the sum rule, and has been
applied in \cite{hyperspl}.

Expanding the sum rule in the parameter $1/m_Q$,
keeping the leading term and the
first order corrections, allows to derive sum rules
for $g$ and for the coefficients
$a$ and $b$ in (\ref{17b}).
In the formula (\ref{split}) for the hyperfine splitting,
only the difference
\be
\Delta_g\equiv g_{P^*P^* \pi}-g_{P^* P \pi}=g \frac{a-b}{m_Q}
\label{u}
\ee
enters. The sum rule for the difference $a-b$ turns out to be quite stable,
 giving \cite{hyperspl}:
\be
a-b \simeq 0.6 \; GeV
\label{ab}
\ee

\resection{Discussion and conclusions}

{}From (\ref{ab}) and
from the formula (\ref{split}) of the hyperfine mass splitting we
obtain:
\be
\Delta_B \approx g^2 (27.3 + 61.4 -75.8)~ MeV= 12.9 g^2 ~MeV
\label{z}
\ee
Notice that we have used in eq. (\ref{split}) $f=f_{\pi}=132 \; MeV$
for all the light pseudoscalar mesons of the octet. This is
suggested by the sum rule for $g$ which shows that $g/f$ is flavour
independent.
 In (\ref{z}) we have  taken $\Lambda_{CSB}=1~GeV$.
 It is evident that there is a large cancellation among the last term and
the other ones. In order to be more quantitative we have to fix the value
of $g$. In Ref. \cite{noi} the range of values
$g \simeq 0.2-0.4$ was found; therefore, putting $g^2=0.1 $,
we would obtain
\be
\Delta_B \simeq 1 ~MeV
\label{z1}
\ee

The application of our results to the
charm case is more doubtful, in view of the large values of the $1/m_c$
correction $(a-b)/m_c$. By scaling the result (\ref{z1}) to the charm case,
one obtains
\be
\Delta_D = \frac{m_b}{m_c}\Delta_B \simeq 4 ~MeV
\label{z2}
\ee

In conclusion, our estimate of $\gis -\gi$ allows to include a previously
neglected term in the loop induced contribution to the hyperfine splitting,
 providing
a substantial cancellation
and reconciliating the chiral calculation with the experimental data.
\par
\vspace*{1cm}
\noindent
{\bf Acknowledgements }\\
 I would like to thank F. Feruglio, R. Gatto and G. Nardulli,
with whom the work
 presented here was done, for the pleasant collaboration and useful
discussions.

\end{document}